\documentstyle[emulateapj]{article}

\righthead{HOGG \& FRUCHTER}
\lefthead{HOSTS OF GAMMA-RAY BURSTS}
\submitted{Accepted for publication in The Astrophysical Journal}

\begin{document}

\title{
	The faint-galaxy hosts of gamma-ray bursts
}
\author{
	David W. Hogg\altaffilmark{1,2} \&
        Andrew S. Fruchter\altaffilmark{3}
}
\altaffiltext{1}{Institute for Advanced Study, Olden Lane, Princeton
       NJ 08540; {\tt hogg@ias.edu}}
\altaffiltext{2}{Hubble Fellow}
\altaffiltext{3}{Space Telescope Science Institute, 3700 San Martin
       Dr, Baltimore MD 21218}

\begin{abstract}
The observed redshifts and magnitudes of the host galaxies of
gamma-ray bursts (GRBs) are compared with the predictions of three
basic GRB models, in which the comoving rate density of GRBs is
(1)~proportional to the cosmic star formation rate density,
(2)~proportional to the total integrated stellar density and
(3)~constant.  All three models make the assumption that at every
epoch the probability of a GRB occuring in a galaxy is proportional to
that galaxy's broad-band luminosity.  No assumption is made that GRBs
are standard candles or even that their luminosity function is narrow.
All three rate density models are consistent with the observed GRB
host galaxies to date, although model (2) is slightly disfavored
relative to the others.  Models (1) and (3) make very similar
predictions for host galaxy magnitude and redshift distributions;
these models will be probably not be distinguished without
measurements of host-galaxy star-formation rates.  The fraction of
host galaxies fainter than 28~mag may constrain the faint end of the
galaxy luminosity function at high redshift, or, if the fraction is
observed to be low, may suggest that the bursters are expelled from
low-luminosity hosts.  In all models, the probability of finding a
$z<0.008$ GRB among a sample of 11 GRBs is less than $10^{-4}$,
strongly suggesting that GRB 980425, if associated with supernova
1998bw, represents a distinct class of GRBs.
\end{abstract}


\section{Introduction}

The study of gamma-ray bursts (GRBs) has been revolutionized by the
discovery of extremely well-localized x-ray, optical and radio
transients (Costa et al 1997, van Paradijs et al 1997, Frail et al
1997).  Follow-up of the optical transients (OTs) has shown that GRBs
come from cosmological distances (Metzger et al 1997b).  One notable
early result of this follow-up is that the OT host galaxies have
generally been near twenty-fifth magnitude in the visible, for those
cases in which a host galaxy has been detected (which is the majority
with OTs; references in Table~\ref{tab:thelist}).  Five of these host
galaxies now have redshifts $z>0.8$ (Metzger et al 1997a, Djorgovski
et al 1998e, Djorgovski et al 1999, Kulkarni et al 1998, Bloom et al
1999).  The question considered here is: {\em How do the flux and
redshift distributions of the GRB hosts compare with the predictions
of simple models?}

It has been surprising to many that bright GRBs, which have a
``euclidean'' number-flux relation suggesting that they are local, are
not correlated on the sky with local, bright galaxies (Schaefer 1998,
Band \& Hartmann 1998).  This lack of association has been termed the
``no-host problem.''  Although authors in the no-host literature have
generally assumed that the GRB luminosity function is narrow, the lack
of correlation at the bright end has suggested, even prior to the
recent redshift determinations, that the typical GRB intrinsic
energies are very great.  The approach taken here is complementary; it
is to compute several different host galaxy flux probability
distribution functions under the simplest possible assumptions about
GRB probability as a function of host galaxy luminosity and redshift,
and detectability as a function of redshift.  The distribution
functions are compared with the observations of GRB hosts.

Unfortunately, the association of each GRB with its host requires
several steps: The burst is first localized to an accuracy of half to
several arcminutes by an x-ray camera.  An OT must be discovered in
the x-ray error box, sometimes with the help of a prior radio
detection.  Finally, the OT must decay sufficiently to allow a search
for an underlying galaxy.  It may not be coincidental that many of the
hosts have $R\approx 25$~mag, comparable to the detection limit of a
few hours' integration on a large telescope.  It is at least possible
in one or two cases that the currently associated galaxy is not the
host but rather a brighter, foreground galaxy at the sky position of
the OT by chance.  This problem may be compounded by the arcsec seeing
in ground-based telescope images.  Despite these caveats, for the
purposes of this work, the conventionally believed associations of
x-ray transients with GRBs, OTs with x-ray transients, and host
galaxies with OTs, will all be accepted with fawning credulity.  As
will, of course, the hypotheses that GRBs are cosmological in origin,
and that they are associated with normal galaxies.

Table~\ref{tab:thelist} lists the GRBs with associated OTs and the
positions and extintinction-corrected magnitudes of their host
galaxies.  Magnitudes have been corrected for extinction, and
redshifts are given where known.  The OT detection associated with GRB
971227 is unconfirmed and its field was therefore not searched
exhaustively for a host galaxy.  This non-detection has little impact
on the results because, as will be shown below, the limiting magnitude
of the search does not put an interesting constraint on its host
galaxy, even if the OT detection is good.  The host galaxies of GRBs
980326 and 980329 have uncertain magnitudes.  The hosts may not be
detected, because the light attributed to the hosts may in fact be
coming partly from the OTs, even in the latest images.  The measured
fluxes really ought to be treated as upper limits (Bloom \& Kulkarni
1998, Fruchter 1999, Pian, private communication).  GRB 980425 is
excluded from the analysis because, as will be shown below, it is an
outlier at the $10^{-4}$ level in all reasonable models.  GRB 980425
must represent a distinct class of bursts.  Finally, the association
of GRB 980613 with its host galaxy is uncertain, because there is a
possibility that no OT was detected at all in this case (Thorsett \&
Metzger, private communication).

An $(\Omega_M,\Omega_{\Lambda})=(0.3,0.0)$ world model is adopted,
except where noted, and all results are independent of the Hubble
constant.  Magnitudes are given in the $R$ band relative to Vega, with
$R=25$~mag corresponding to $R_{\rm AB}=25.25$~mag or
$f_{\nu}=0.29~{\rm \mu Jy}$.

\section{Fiducial models}

Our procedure is to compare several fiducial models with the data
and then discuss the effect of variations in these models on the
comparison.  The emphasis is on minimizing the total number of
assumptions.

In all models, we assume that the probability of a GRB ``going off''
in a particular galaxy, at a particular epoch, is proportional to that
galaxy's broad-band luminosity.  In the star-formation-rate (SFR)
models, it is assumed that the total comoving rate density (number per
unit comoving volume per unit time) at which GRBs are produced at any
particular epoch is proportional to the total comoving star formation
rate density $\dot\rho(z)$ at that epoch.  A by-eye fit was performed
to the $\dot\rho(z)$ measurements of Connolly et al (1997); this fit
is shown in the second panel of Figure~\ref{fig:inputs}, it has
$\dot\rho(z)\propto z^{0.90}$ at redshifts $z<1.0$, $z^{0.00}$ at
$1.0<z<2.5$, and $z^{-0.38}$ at $z>2.5$.  In the total-stellar-density
(TSD) models, it is assumed that the comoving rate density is
proportional to the total number density of stars which have been
formed since the beginning of cosmic time, the integral of the star
formation rate density $\int\dot\rho(z)\,dt$.  Finally, in the
constant per comoving volume (CCV) model, the comoving rate density is
the same at all epochs.

At least some GRBs and OTs can be detected to very high redshift
(Kulkarni et al 1998).  Unfortunately, the detection function $p_{\rm
detect}(z)$, or probability of GRB (and X-ray and OT) detection as a
function of redshift $z$, is unknown empirically and impossible to
compute theoretically because it depends not only on the sensitivities
of the detectors but on the distribution of intrinsic gamma-ray, x-ray
and optical properties of the bursts, along with the quality and
consistency of x-ray and optical follow-up observations.  Although it
is somewhat unconventional to pack all of the uncertainties about the
multivariate gamma-ray, x-ray, and optical GRB luminosity functions
and detector and follow-up sensitivities into the single function
$p_{\rm detect}(z)$, it greatly reduces the total number of
assumptions and clarifies the model-dependence of the results.
Studies of the GRB luminosity function which are consistent with the
observed GRB number counts and redshifts suggest that $p_{\rm
detect}(z)$ is a weak function of $z$, falling by only a factor of a
few from $z=1$ to $z=3$ (Krumholz, Thorsett \& Harrison 1998).
Indeed, at least one burst with an associated OT has been associated
with a redshift 3.4 host galaxy (Kulkarni et al 1998), and it is
plausible that GRB 980329 is at $z\sim 5$ (Fruchter 1999).  In any
event, as discussed below, we find that the results are only
significantly affected if $p_{\rm detect}(z)$ is very strongly
weighted towards low redshift (corresponding to a GRB or X-ray or OT
luminosity function very strongly weighted towards low-energy bursts).
For the purposes of the fiducial models it is simply assumed that
$p_{\rm detect}(z)\propto (1+z)^{-1}$ over the redshift range $0<z<5$
and $p_{\rm detect}(z)=0$ at $z>5$, as shown in the top panel of
Figure~\ref{fig:inputs}.  The function $p_{\rm detect}(z)$ varies
slowly out to $z=5$ because the gamma-ray bursts are not assumed to be
standard candles; this analysis allows the luminosity function to be
very wide without in fact specifying its width or shape.  At $z>5$
$p_{\rm detect}(z)$ vanishes because Lyman limit absorption will
obscure OTs and host galaxies in the $R$ band.  As will be seen below,
the results do not depend very strongly on the assumed form of $p_{\rm
detect}(z)$.

We assume that all observations of hosts are performed in the R band.
Thus the observing band in the frame of the host will vary with
redshift.  To maintain independence of world model, the characteristic
luminosity $L^{\ast}$ appearing in the Schechter (1976) form of the
luminosity function is input in the form of the apparent magnitude
$R^{\ast}$ to which it corresponds at each epoch, which is the
directly observed quantity.  In practice, the $R^{\ast}(z)$ employed
is equivalent to $\log L^{\ast}$ evolving from $36.5$ (in $\nu
L_{\nu}$ in $h^{-2}\,{\rm W}$) at $z=0$ to $40.0$ at $z=5$ in an
$(\Omega_M,\Omega_{\Lambda})=(0.1,0.0)$ universe.  (This is not the
default cosmology, but this form for $R^{\ast}(z)$ is simply a
parameterization of the observational determinations, with the
$(0.1,0.0)$ world model used for consistency with Pozetti et al 1998).
This form of $R^{\ast}(z)$ is shown in Figure~\ref{fig:inputs} and is
consistent with all measures of luminosity function evolution to
$z\sim 1$ (Lilly et al 1995, Ellis et al 1996, Hogg 1998) and at
$z>2.5$ (Pozzetti et al 1998).  As shown in Figure~\ref{fig:inputs},
the faint-end slope parameter $\alpha(z)$ is chosen to be flat
($\alpha=-1.00$) in the local Universe (eg, Loveday et al 1992, Lilly
et al 1995, Ellis et al 1996, Hogg 1998) and slightly steeper
($\alpha=-1.30$) at high redshifts $2.5<z<5$ (Pozzetti et al 1998) and
steeper still ($\alpha=-1.75$) in between at redshifts $0.6<z<2.0$.
This $\alpha=-1.75$ epoch is required to make the steep number counts,
which show $d\log N/dm=0.3$ in the $R$ band at the faint end (Hogg et
al 1997), and in the redshift interval $0.6<z<2.0$ there are not yet
strong direct constraints on this slope (Lilly et al 1995, Ellis et al
1996, Hogg 1998), so this $\alpha(z)$ model, shown in
Figure~\ref{fig:inputs}, is consistent with all observations.  This
model is only arbitrary in the choice of redshift interval for the
$\alpha=-1.75$ epoch; some such epoch is required in all natural
models of the faint galaxy counts.

In the SFR models, the GRB probability will not be strictly
proportional to a galaxy's broad-band luminosity but rather to its
star formation rate.  At high redshift, the observed visual luminosity
is a very good measure of star formation rate, because observed visual
is emitted in the rest-frame ultraviolet.  This is less true in the
local Universe where star formation is at least somewhat weighted
towards lower-luminosity galaxies (Small et al 1997).  This effect is
not strong and therefore does not greatly affect the results but means
that the number of bright ($R<22$) hosts predicted by this procedure
may be slightly higher than in a more accurate representation of the
SFR model.

\section{Fiducial results and comparison with data}

The host galaxy flux and redshift distribution predictions of the
fiducial models are shown in Figure~\ref{fig:fluxdist}, along with the
observed host galaxy magnitudes and redshifts.

The models are compared with relative likelihoods $\cal L$ of
obtaining the observed host galaxy magnitudes given the model.  The
likelihoods are computed by multiplying together the differential
probability $f(m)$ (probability per unit magnitude) evaluated at each
observed host magnitude value, and the integral of $f(m)\,dm$ from
$m_{\rm lim}$ to $\infty$ for the magnitude limits on GRBs 971227,
980326 and 980329.  Unfortunately, likelihoods are only relative, not
absolute.  The relative likelihoods for the fiducial SFR,TSD,CCV
models are $1.00:0.12:0.57$.

Although clearly all three models are consistent with the observed
host-galaxy magnitude distribution, both the likelihoods and the
appearence of Figure~\ref{fig:fluxdist} suggests that the TSD model is
slightly disfavored relative to SFR and CCV.  The likelihood test is
not applied to the redshift distribution because there is great
uncertainty in the redshift identification probability as a function
of magnitude and redshift, which may dominate the shape of the
observed redshift distribution.  However, under the assumptions about
redshift identification probability with which
Figure~\ref{fig:fluxdist} was made, it appears that the host-galaxy
redshift distribution also slightly disfavors the TSD model.

In all three models, the probability of finding a $z<0.008$ GRB among
a sample of 11 GRBs is less than $10^{-4}$, strongly suggesting that
GRB 980425, if associated with supernova 1998bw, represents a distinct
class of GRBs, and justifying its exclusion from the analysis.  It is
worthy of note that this argument for a second class of GRBs makes no
reference to the intrinsic energetics of GRB 980425 and is therefore
qualitatively different from previous arguments (Kulkarni et al
1998b).

It is clear from Figure~\ref{fig:fluxdist} that even with much larger
number of GRB host observations it will be very difficult to
distinguish the SFR from the CCV using the magnitude or redshift
distributions.  Previous claims to the contrary (Totani 1998) are
based on an unrealistic assumption that GRBs are close to standard
candles.  The SFR and CCV models make very similar predictions because
the comoving rate densities only differ significantly at low redshift,
where there is not much comoving volume, and at high redshift, where
the time dilation $(1+z)$ factor which comes into rate calculations
and the declining $p_{\rm detect}(z)$ both effectively reduce the
contribution to the total GRB rate.  The two hypotheses will be
readily distinguishable by investigating the spectral properties of
the associated hosts; the SFR models predict bluer and more
emission-line-dominated galaxies than an average sample.  It does
appear that the majority of GRB hosts do show signs of fairly active
star formation (Kulkarni et al 1998, Metzger et al 1997a, Fruchter et
al 1998); there may already be enough information about host galaxies
to distinguish these models.  Another simple hypothesis which would
make very similar predictions to the SFR is that the comoving rate
density is proportional to the evolving number density of quasars (eg,
Schmidt, Schneider \& Gunn 1995).

Previous no-host studies have claimed to rule out interesting GRB
models with limits on host galaxies in the range 13 to 23~mag
(Schaefer 1998, Band \& Hartmann 1998), but such studies do not
strongly constrain the GRB models presented here.  There may be no
contradiction, because the previous literature on the no-host problem
is primarily concerned with very bright bursts, and, a narrow or
standard-candle GRB luminosity function usually has been assumed.  The
present analysis, which does not specify a GRB luminosity function but
allows it to be very wide, is not capable of making different
predictions for the host galaxies of bursts with different observed
fluences.  This analysis sacrifices that capability in order to avoid
making unnecessary assumptions.

\section{Variation with inputs}

Not surprisingly, the predictions do not depend strongly on cosmology.
In an $(\Omega_M,\Omega_{\Lambda})=(1.0,0.0)$ universe, the
SFR,TSD,CCV models have likelihoods $1.00:0.06:0.45$.  In an
$(\Omega_M,\Omega_{\Lambda})=(0.4,0.6)$ universe, they have
$1.00:0.12:0.57$.

Unfortunately, the results do depend somewhat on the choice of $p_{\rm
detect}$, the least well-constrained of the model inputs.  If $p_{\rm
detect}(z)=1$ is adopted, the SFR,TSD,CCV models have likelihoods
$1.00:0.23:0.58$.  If $p_{\rm detect}(z)=(1+z)^{-3}$ is adopted, they
have $1.00:0.03:0.23$.  Weighting $p_{\rm detect}(z)$ towards high
redshift improves the success of the TSD model relative to the SFR and
CCV models, because the TSD rate density itself is weighted towards
low redshift.  However, $p_{\rm detect}(z)=1$ is clearly an
unrealistic model; it says that GRBs (and x-ray transients and OTs)
are equally easy to detect at all redshifts!  Conversely, strong
weighting of $p_{\rm detect}(z)$ towards low redshift improves the
success of SFR and CCV relative to TSD.  Even in the $p_{\rm
detect}(z)=(1+z)^{-3}$ models, the probability of finding a $z<0.008$
GRB among this sample of 11 is still very small.

There is some debate about the rise of the star formation rate density
with cosmic time at high redshift, since the measurements are subject
to possible incompleteness and uncertain dust extinction corrections
(eg, Pettini et al 1998).  This uncertainty is not important here; if
the rise in the star formation rate with time at $z>2.5$ is replaced
with a constant value equal to the value at $z>1.0$ (which may more
accurately represent the true situation, Steidel \& Adelberger,
private communication), the likelihoods for the SFR,TSD,CCV models
become $1.00:0.20:0.64$.  It is worthy of note that this change makes
the SFR and CCV models even more difficult to distinguish than in the
fiducial case.

The uncertain high-redshift faint-end slope of the galaxy luminosity
function does affect the results.  If it is changed to
$\alpha(z)=-1.75$ for all redshifts $z>2.0$, which is probably still
consistent with the existing $z\sim 3$ galaxy observations (Pozzetti
et al 1998), the SFR,TSD,CCV models have likelihoods $1.00:0.29:0.41$.
The relative success of TSD is improved when the luminosity function
is made more dwarf-rich.  However, the fraction $F_{>28}$ of hosts
predicted to be at $R>28$~mag (ie, extremely faint) becomes large.
Quantitatively, as the $z>2.5$ value of $\alpha$ ranges from $-1.0$ to
$-1.75$, $F_{>28}$ for the SFR,TSD,CCV models ranges from $F_{>28}=
0.21,0.19,0.17$ to $0.34,0.22,0.35$.  If the faint end of the galaxy
luminosity function really is steep at high redshift, and either the
SFR or the CCV hypothesis is close to correct, it is possible that
some of the current GRB host galaxy identifications or photometric
measurements are in error, since very few optical observations are
sensitive to 28~mag.

In all these models, a large fraction of host galaxies have extremely
small intrinsic luminosities.  Even if such galaxies are as common as
the extrapolated galaxy luminosity functions suggest, they may not
host GRBs.  For example, in neutron-star--neutron-star merger
scenarios, very low-mass galaxies do not gravitationally bind kicked
neutron-star binaries (Bloom, Sigurdsson \& Pols 1998a).  For this
reason it might be sensible to implement a low-luminosity cutoff to
the galaxy luminosity function.  With a cutoff at $10^{-2}\,L^{\ast}$
the SFR,TSD,CCV models have likelihoods $1.00:0.04:0.58$, and
$F_{>28}$ drops to 0.030,0.0068,0.044.  Furthermore, the dependence of
the results on the faint-end slope of the luminosity function at high
redshift disappears almost entirely.

\section{Summary}

The expected distribution of GRB host galaxy fluxes and redshifts are
predicted, assuming reasonable GRB comoving rate density models and
that at any epoch, GRB probability is proportional to host galaxy
luminosity.  The analysis makes fewer assumptions than previous
studies.  In particular, it makes no assumption of a narrow GRB
luminosity function.  The agreement between the models and the data,
for reasonable choices of model parameters, is very good.  Of the
three models considered, the TSD fares worst, although it is by no
means ruled out.  The SFR and CCV models make very similar predictions
for the host galaxy magnitude and redshift distributions, so those two
models will have to be distinguished with observations of host-galaxy
star-formation rates.

We do not find a classical no-host problem, in the sense of a lack of
local, bright galaxy hosts, although as stated above, the present
analysis does not make different predictions for bursts of different
fluences.  There may be some suggestion that GRBs do not occur in
extremely low-luminosity host galaxies, because, when no cut-off is
applied to the luminosity function at low luminosity, the models
predict a significant fraction of GRB hosts below the detection limits
of typical surveys.  Of course it is possible that up to three of the
current hosts fall into this category in the current sample of GRBs
with OTs.

One conclusion of this work is that GRB 980425, associated with the
low-redshift supernova 1998bw, must be a member of a distinct class.
In all models, the probability of finding a $z<0.008$ GRB among a
sample of 11 GRBs is less than $10^{-4}$.

It is notable that in the models presented here, many GRBs and their
hosts lie in the redshift range $1.3<z<2.5$, where galaxies are very
hard to identify with visual spectroscopy, even on large telescopes.
Either infrared spectroscopy or the ultraviolet capabilities of the
Hubble Space Telescope may be necessary to obtain the redshifts of
these GRBs.

{\em A note on history: The first version of this paper was submitted
when all known GRB host magnitudes (except GRB 980425) were in the
range $24.4<R<25.8$~mag.  At that time, the largest discrepancy
between the observations and the models was that the width of the
observed magnitude distribution was much narrower than the prediction
of any model.  Since then, three host magnitude measurements (971214,
980326, and 980613) have been significantly revised (Odewahn et al
1998, Bloom \& Kulkarni 1998, Metzger et al private communication),
and two new host magnitudes (980703 and 990123) have been measured
(Bloom et al 1998c, Fruchter et al 1999), greatly improving the
agreement between the models and the data.}

\acknowledgements We thank John Bahcall, George Djorgovski, Fiona
Harrison, Shri Kulkarni, Steve Thorsett and Eli Waxman for useful
discussions, Holger Pedersen for results in advance of publication,
and Jochen Greiner for maintaining his comprehensive GRB website.
Support for DWH was provided by Hubble Fellowship grant
HF-01093.01-97A from STScI, which is operated by AURA under NASA
contract NAS~5-26555.


\begin{deluxetable}{rrrrrrrrl}
\tablewidth{0pt}
\scriptsize
\tablecaption{
        Host galaxy information for GRBs with associated OTs\label{tab:thelist}
}
\tablehead{
   \colhead{GRB}
 & \colhead{RA (2000)}
 & \colhead{Dec (2000)}
 & \colhead{$l$}
 & \colhead{$b$}
 & \colhead{$A_R$\tablenotemark{a}}
 & \colhead{$R_{\rm corr}$}
 & \colhead{$z$}
 & \colhead{references}
\\
   \colhead{}
 & \colhead{(h\,m\,s)}
 & \colhead{($^{\circ}\,'\,''$)}
 & \colhead{(deg)}
 & \colhead{(deg)}
 & \colhead{(mag)}
 & \colhead{(mag)}
 & \colhead{}
 & \colhead{}
}
\startdata
970228                  & $05~01~46.7$ & $+11~46~53.6$ &
  $188.91$ & $-17.94$ & $0.58$ & $24.7$ &         &
  Fruchter et al (1998) \nl
970508                  & $06~53~49.4$ & $+79~16~19.6$ &
  $134.96$ & $+26.73$ & $0.13$ & $25.0$ & $0.835$ &
  Bloom et al (1998b), Metzger et al (1997a) \nl
971214                  & $11~56~26.0$ & $+65~12~00.0$ &
  $132.04$ & $+50.94$ & $0.04$ & $26.2$ & $3.418$ &
  Kulkarni et al (1998a), Odewahn et al (1998) \nl
971227\tablenotemark{b} & $12~57~10.6$ & $+59~24~43.0$ &
  $121.57$ & $+57.70$ &        & $>22.0$ &         &
  Mendez, Ruiz-Lapuente \& Walton (1998) \nl
980326                  & $08~36~34.0$ & $-18~51~24.0$ &
  $242.36$ & $+13.04$ & $0.22$ & $>27.1$ &         &
  Bloom \& Kulkarni (1998) \nl
980329\tablenotemark{c} & $07~02~38.0$ & $+38~50~44.0$ &
  $178.12$ & $+18.65$ & $0.19$ & $>25.5$ &         &
  Djorgovski et al (1998c) \nl
980425\tablenotemark{d} & $19~35~03.2$ & $-52~50~46.1$ &
  $344.99$ & $-27.72$ & $0.16$ & $14.1$ & $0.008$ &
  Kulkarni et al (1998b) \nl
980519                  & $23~22~21.4$ & $+77~15~43.0$ &
  $117.96$ & $+15.26$ & $0.71$ & $24.8$ &         &
  H. Pedersen, private communication \nl
980613\tablenotemark{e} & $10~17~57.6$ & $+71~27~26.4$ &
  $138.06$ & $+40.86$ & $0.23$ & $23.4$ & $1.096$ &
  Metzger, private communication, Djorgovski et al (1999) \nl
980703                  & $23~59~06.7$ & $+08~35~07.0$ &
  $101.48$ & $-52.26$ & $0.15$ & $22.6$ & $0.966$ &
  Bloom et al (1998c), Djorgovski et al (1998e) \nl
990123                  & $15~25~30.5$ & $+44~46~00.5$ &
   $73.12$ & $+54.64$ & $0.04$ & $23.7$ & $1.600$\tablenotemark{f} &
  Fruchter et al (1999), Bloom et al (1999) \nl
\enddata
\tablenotetext{a}{Extinction values for the $R$ band are based on the
reddening maps of Schlegel, Finkbeiner \& Davis (1998).}
\tablenotetext{b}{Included in analysis although OT detection is
uncertain and unconfirmed.}
\tablenotetext{c}{Host galaxy magnitude is considered a limit because
it was measured by subtracting extrapolation of fading OT flux; for
the purposes of the comparing models and observations, $25.5$~mag is
adopted in this study.}
\tablenotetext{d}{Excluded from analysis because this burst must come
from a distinct class; see text.}
\tablenotetext{e}{Included in analysis although OT detection is
uncertain (Thorsett \& Metzger, private communication).}
\tablenotetext{f}{Redshift is uncertain because it is based only on
a strong absorption system in the spectrum of the OT.}
\end{deluxetable}

\clearpage
\plotone{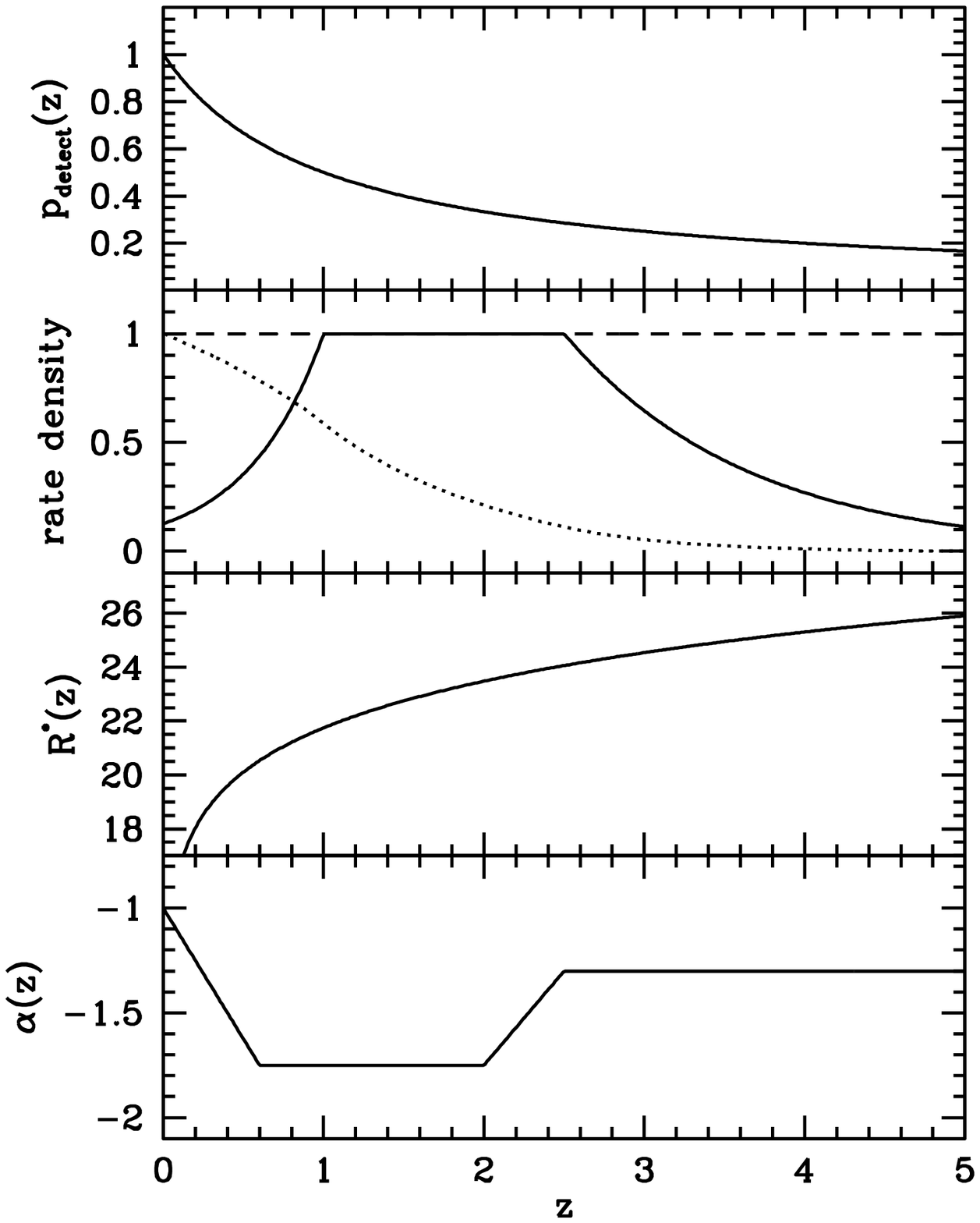}
\figcaption[inputs.eps]{The inputs to the fiducial models: {\sl (top)}
the detection function $p_{\rm detect}(z)$ as a function of redshift
$z$, {\sl (second)} the comoving rate density for the SFR model {\sl
(solid)}, TSD model {\sl (dotted)} and CCV model {\sl (dashed)}, {\sl
(third)} the apparent magnitude $R^{\ast}(z)$ corresponding to
$L^{\ast}$ in the observed $R$ band, and {\sl (bottom)} the faint-end
slope $\alpha(z)$ of the galaxy luminosity function, appropriate for
the observed $R$ band.\label{fig:inputs}}

\clearpage
\plotone{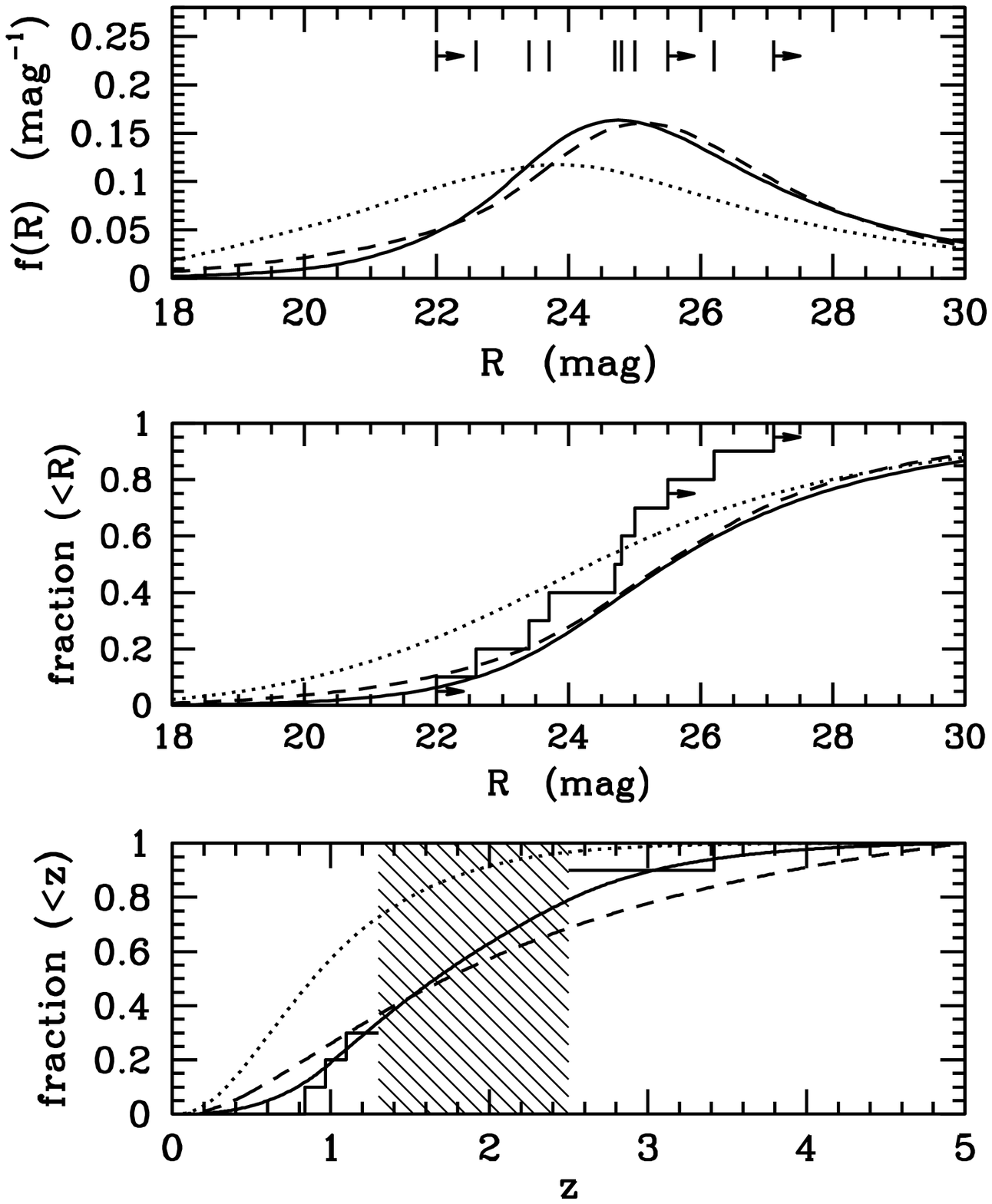} \figcaption[fluxdist.eps]{The differential
distribution of host galaxy magnitudes $R$ {\sl (top)}, cumulative
distributions of host galaxy magnitudes {\sl (middle)}, and cumulative
distribution of host galaxy redshifts $z$ {\sl (bottom)}, for the
fiducial SFR model {\sl (solid)}, TSD model {\sl (dotted)} and CCV
model {\sl (dashed)}.  Vertical bars show the observed host galaxy
magnitudes, histograms show the observed cumulative magnitude and
redshift distributions.  Limits are marked with arrows.  The plotted
magnitudes have been corrected for extinction.  In the redshift plot,
it is assumed that all hosts with no redshift lie in the shaded
redshift range $1.3<z<2.5$, because visual spectroscopy is difficult
between the redshift at which the [O\,II] 3727\,\AA\ line leaves the
red end of the spectroscopic window that that at which the Ly$\alpha$
1216\,\AA\ line enters the blue.  (Note that spectroscopy with large
telescopes has been performed on most of the known GRB host galaxies.)
The redshift of GRB 990123 is not plotted because it is based on an
absorption system in the OT and is therefore both uncertain and not
subject to the same selection effects as the emission
redshifts.\label{fig:fluxdist}}

\end{document}